\documentclass{elsart}
\usepackage{natbib}
\usepackage{graphicx}
\usepackage{amssymb}

\begin{document}

\begin{frontmatter}

\title{Theory of weakly damped free-surface flows: a new formulation based on potential
flow solutions}
%
%
\author[CMLA]{F. Dias \corauthref{cor}}
\ead{Frederic.Dias@cmla.ens-cachan.fr}
\corauth[cor]{Corresponding author.}
\author[Landau]{A.I. Dyachenko}
\ead{alexd@landau.ac.ru}
\author[Tucson,Landau,Phian]{V.E. Zakharov\thanksref{also2}}
{
\thanks[also2]{Waves and Solitons LLC, W. Sereno Dr., Gilbert, AZ, 85233, USA} 
\ead{zakharov@math.arizona.edu}
}

\address[CMLA]{CMLA, ENS Cachan, CNRS,
  PRES UniverSud, 61 Av. President Wilson,
  F-94230 Cachan, FRANCE}
\address[Landau]{Landau Institute for Theoretical Physics, 
2 Kosygin str., Moscow, 119334, Russia}
\address[Tucson]{Department of Mathematics, University of
Arizona, Tucson, AZ, 857201, USA}
\address[Phian]{Lebedev Physical Institute RAS, 
53 Leninskii pr., Moscow, 119991, Russia}

\begin{abstract}
Several theories for weakly damped free-surface flows have been formulated. In this paper we
use the linear approximation to the Navier-Stokes equations to derive a new set of equations for potential flow which
include dissipation due to viscosity. A viscous correction is added not only to the irrotational pressure (Bernoulli's
equation), but also to the kinematic boundary condition. The nonlinear Schr\"odinger (NLS) equation that one can derive
from the new set of equations to describe the modulations of weakly nonlinear, weakly damped deep-water gravity waves turns out
to be the classical damped version of the NLS equation that has been used by many authors without rigorous justification.
\end{abstract}

\end{frontmatter}

\section{Introduction}

Even though the irrotational theory of free-surface flows can predict successfully many observed wave 
phenomena, viscous effects cannot be neglected under certain circumstances. Indeed the question of dissipation in potential flows 
of fluid with a free surface is a very important one. As stated by \cite{LH1992},
it would be convenient to have equations and boundary conditions of comparable simplicity as for undamped
free-surface flows. The peculiarity here lies in the fact that the viscous term in the Navier-Stokes equations is identically 
equal to zero for a velocity deriving from a potential.

The effect of viscosity on free oscillatory waves on deep water was studied by \cite{Boussinesq} and \cite{Lamb1932}, among
others. \cite{Basset} also worked on viscous damping of water waves. It should be pointed out that the famous treatise on hydrodynamics 
by Lamb has six editions. The paragraphs on wave
damping are not present in the first edition (1879) while they are present in the third edition (1906). The authors did not
have access to the second edition, so it is possible that Boussinesq and Lamb published similar results at the same time.
Lamb derived the decay rate of the linear wave amplitude in two different ways: in \S\,348 of the sixth edition by a dissipation 
calculation (this is also what Boussinesq did) and in \S\,349 by a direct calculation based on the linearized Navier-Stokes equations. 
Let $\alpha$ denote the wave amplitude, $\nu$ the kinematic viscosity of the fluid and $k$ the wavenumber of the decaying wave. 
Lamb showed that
\begin{equation}\label{Lamb_decay}
\frac{\partial\alpha}{\partial t} = -2\nu k^2 \alpha. 
\end{equation}
Equation (\ref{Lamb_decay}) leads to the classical law for viscous decay of waves of amplitude $\alpha$, namely
$\alpha \sim \exp(-2\nu k^2 t)$.

In order to include dissipation accurately into potential flow solutions, one must somehow take into account vorticity.
In the papers by \cite{JosephWang2004}, \cite{WangJoseph2006}, vorticity was taken into account only in Bernoulli's
equation, while only the potential component of the velocity was used in the kinematic condition. \cite{Tuck},
as is commonly done in the ship research community, derived a single linearized free-surface condition with a
dissipative term, that combines the kinematic and dynamic boundary conditions. \cite{RFF1991} added the
vortical component in the kinematic boundary condition but did not simplify it. \cite{LH1992} simplified the equations
of \cite{RFF1991} by introducing a new free surface differing from the classical free surface by the integral over
time of the vortical component of the velocity. Below we proceed differently and show that the
small vortical component of the velocity plays a role in the kinematic condition as a dissipative term. The new resulting
set of equations, which describes potential flow with dissipation, leads to a dispersion relation which corresponds
exactly to that of \cite{Lamb1932} in the limit of small viscosity.

\section{Derivation of the new set of equations in the linear approximation}

In order to clearly show how the new set of equations is derived, we first introduce the correction due to viscosity in the 
linearized equations for the potential flow of an incompressible fluid with a free surface 
(the linearization applies to the dynamic boundary condition expressed through Bernoulli's equation and the kinematic boundary 
condition on the free surface). We consider an ``almost'' potential flow, with its velocity $\vec v$ given by
\begin{equation}\label{N-S_velocity}
\vec v = \nabla\phi + \nabla \times \vec A, 
\end{equation}
where $\vec A$ is a vector stream function. It is assumed that the vortical component of the velocity $\nabla \times \vec A$ 
is much less than the potential component of the velocity $\nabla\phi$.

It is important to emphasize that we can analyse the correction due to dissipation even in the linear approximation to
the Navier-Stokes equations. Therefore we consider a linear and ``almost'' potential flow in the two-dimensional (2D) case.
Assuming for simplicity that the water depth is infinite, the equations for the velocity $\vec v=(u,w)$ read
\begin{eqnarray}\label{linear_N-S}
\frac{\partial u}{\partial t} &=& -\frac{1}{\rho} \frac{\partial p}{\partial x} 
		+\nu \left ( \frac{\partial^2 u}{\partial x^2} + \frac{\partial^2 u}{\partial z^2} \right ), \cr
\frac{\partial w}{\partial t} &=& - g -\frac{1}{\rho} \frac{\partial p}{\partial z} 
		+\nu \left ( \frac{\partial^2 w}{\partial x^2} + \frac{\partial^2 w}{\partial z^2} \right ),
\end{eqnarray}
with the condition of flow incompressibility
\begin{equation}\label{incompressibility}
\frac{\partial u}{\partial x} + \frac{\partial w}{\partial z} = 0. 
\end{equation}
Using the Helmholtz decomposition for the velocity (\ref{N-S_velocity}) yields
\begin{eqnarray}\label{Helmholtz}
u(x,z,t) &=& \frac{\partial \phi}{\partial x} - \frac{\partial A_y}{\partial z},\cr
w(x,z,t) &=& \frac{\partial \phi}{\partial z} + \frac{\partial A_y}{\partial x},
\end{eqnarray}
since in 2D there is a single component to the vector stream function, which we denote by $A_y$.
Equations (\ref{linear_N-S}) combined with equation (\ref{incompressibility}) can be written as follows:
\begin{eqnarray}\label{linear_solution}
\frac{\partial A_y}{\partial t} &=& \nu \left ( \frac{\partial^2 A_y}{\partial x^2} + \frac{\partial^2 A_y}{\partial z^2} \right ),\cr
\frac{\partial\phi}{\partial t} &=& -\frac{p(x,z,t)}{\rho} - g z 
+ \frac{p_0}{\rho}.
\end{eqnarray}
To determine the `normal modes' which are periodic in respect of $x$ with a prescribed wavelength $2\pi/k$, we assume
a time-factor $e^{-i\omega t}$ and a space-factor $e^{ikx}$. The solutions for
the potential $\phi$ and the single component of the vector potential $A_y$ are then
\begin{eqnarray}\label{potentialS}
\phi(x,z,t) &=& \phi_0 e^{i(kx-\omega t)} e^{|k| z},\cr
{A_y}(x,z,t) &=& {A_y}_0 e^{i(kx-\omega t)} e^{m z},
\end{eqnarray}
where 
\begin{equation}\label{m_and_k}
m^2 = k^2 -i\frac{\omega}{\nu}.
\end{equation}
Equation (\ref{m_and_k}) can be rewritten as
\begin{equation}\nonumber
m = \frac{1}{\sqrt{2}}\left ( \sqrt{\sqrt{k^4+\frac{\omega^2}{\nu^2}}+k^2} - i\sqrt{\sqrt{k^4+\frac{\omega^2}{\nu^2}}-k^2}\right ).
\end{equation}
Note that in the inviscid limit, both the vector stream function $A$ and the viscosity $\nu$ are equal to zero.

Let us now write down the boundary conditions along the free surface.
The linearized kinematic boundary condition reads
\begin{equation}\label{linear_kinematic_N_S}
\frac{\partial \eta}{\partial t} = w(x,0,t), 
\end{equation}
since the velocity is evaluated at $z=0$ in the linear approximation. 
It yields
\begin{equation}\label{linear_eta}
\eta(x,t) = \frac{1}{\omega}\left(i|k|\phi_0 - k{A_y}_0\right)e^{i(kx-\omega t)}.
\end{equation}
The linearized dynamic boundary conditions read
\begin{eqnarray}\label{linear_Dynamic_N_S}
p - 2\rho\nu \frac{\partial w}{\partial z} &=& p_0 \hspace{1cm}\mbox{at $z=\eta(x,t)$},\cr
\nu \left ( \frac{\partial u}{\partial z} + \frac{\partial w}{\partial x} \right ) &=& 0 \hspace{1cm}\mbox{at $z=0$}.
\end{eqnarray}
The boundary conditions (\ref{linear_kinematic_N_S}) and (\ref{linear_Dynamic_N_S}) provide two pieces of information:
\begin{enumerate}
\item the relationship between the potential $\phi_0$ and the vector stream function ${A_y}_0$,
\begin{equation}\label{phi_and_A}
{A_y}_0 = \frac{2 i |k|k}{m^2 + k^2}\phi_0 = -2 \left(\frac{\nu}{\omega}\right)\frac{|k|k\phi_0}{1 + 2i k^2 \frac{\nu}{\omega}},
\end{equation}
\item the dispersion relation $\omega(k)$
\begin{equation}\label{Dispersion_relation_N_S}
\left ( 2 - \frac{i\omega}{\nu k^2} \right )^2 + \frac{g}{\nu^2 |k|^3} = 
4\left ( 1 - \frac{i\omega}{\nu k^2} \right )^{1 \over 2}.
\end{equation}
\end{enumerate}
In the limit of small viscosity $\nu k^2 \ll \omega$, the right-hand side of
(\ref{Dispersion_relation_N_S}) vanishes and the following approximation for $\omega$ can be derived:
\begin{equation}\label{Dispersion_relation_N_S_approx}
\omega = \pm\sqrt{g |k|} -2i\nu k^2.
\end{equation}
Equation (\ref{Dispersion_relation_N_S_approx}) leads to the classical law (\ref{Lamb_decay}) for viscous decay of free waves. 
Under the same limit ($\nu k^2 \ll \omega$), equation (\ref{phi_and_A})
can be written as follows:
\begin{equation}\label{phi_and_A_xz}
\frac{\partial^2 {A_y}}{\partial x\partial t} = 
2\nu \frac{\partial^3\phi}{\partial x^2\partial z}.
\end{equation}
This equation was derived by \cite{RFF1991}.

So far, we have only described the viscous solution to the linear Navier-Stokes equations with small viscosity and as said
in the introduction it is a well-known solution. But at this point a natural question arises: 
{\it Can we describe this flow by using only the potential part of the velocity?}

First, let us look at the kinematic condition (\ref{linear_kinematic_N_S}) and explicitly separate the potential and vortical
components of the velocities in it:
\begin{equation}\label{linear_kinematic_N_S_sep}
\frac{\partial \eta}{\partial t} = 
\frac{\partial \phi}{\partial z} + \frac{\partial A_y}{\partial x} \hspace{1cm}\mbox{at $z=0$.} 
\end{equation}
The vortical part of the velocity ${\partial A_y}/{\partial x}$ can be written in different ways, but here
we want to express it in terms of $\eta$. (Note that we consider linear equations and can therefore freely 
express one function through another.) To do this, one can use the solution for $\eta$ (\ref{linear_eta}) with the
relation (\ref{phi_and_A}) between the constants $\phi_0$ and ${A_y}_0$. Thus, the following relation can be easily
derived:
\begin{equation}\label{damp_1}
\frac{\partial A_y}{\partial x} = -2i|k|^3\left(\frac{\nu}{\omega}\right)\frac{\phi_0}{1+2ik^2\frac{\nu}{\omega}}e^{i(kx-\omega t)}
= -2k^2\nu\eta(x,t) = 2\nu\frac{\partial^2\eta}{\partial x^2}.
\end{equation}
These equations are one of the main results of our paper. 
The kinematic boundary condition now can be written without the vortical component of the velocity: 
\begin{equation}\label{linear_kinematic_N_S_new}
\frac{\partial \eta}{\partial t} = 
\frac{\partial \phi}{\partial z} + 2\nu\frac{\partial^2\eta}{\partial x^2} \hspace{1cm}\mbox{at $z=0$.} 
\end{equation}

Let us now turn to the equation for the potential component of the velocity. The dissipative correction to this equation was discussed 
in \cite{JosephWang2004}, \cite{WangJoseph2006}. It is simply Bernoulli's equation, except that the
pressure must be replaced by 
\begin{equation}\label{new_pressure}
p = 2\rho\nu \frac{\partial w}{\partial z} + p_0 = p_0 + 2\rho\nu
\left ( \frac{\partial^2 \phi}{\partial z^2} + \frac{\partial^2 A_y}{\partial x \partial z} \right ).
\end{equation}
Note that the vortical component in the pressure (\ref{new_pressure}) is of order $\nu^2$ and can be neglected.
So, we can write down Bernoulli's equation with dissipation
\begin{equation}\label{linear_Bernoulli}
\frac{\partial \phi}{\partial t} + g\eta = 
-2\nu \frac{\partial^2 \phi}{\partial z^2} \hspace{1cm}\mbox{at $z=0$.}
\end{equation}
It is easy to check that the two boundary conditions (\ref{linear_kinematic_N_S_new}) and (\ref{linear_Bernoulli})
lead to the same dispersion relation as (\ref{Dispersion_relation_N_S_approx}).

\section{Fully nonlinear equations with dissipation}

In the previous section, we added viscous terms in the {\it linear} equations of the 2D potential flow of a fluid with a free surface.
It is clear that these terms can be added in the {\it nonlinear} equations for potential flow. Moreover the analysis can be extended 
to three-dimensional flows:
\begin{eqnarray}\label{B_E_damped}
\frac{\partial\phi}{\partial t} + \frac{1}{2}|\nabla \phi|^2 + 
g\eta &=& - 2\nu\frac{\partial^2 \phi}{\partial z^2} \hspace{1cm}\mbox{at $z=0$,}\cr
      \frac{\partial\eta}{\partial t} + \nabla\eta \cdot \nabla\phi &=& \frac{\partial\phi}{\partial z} 
+ 2\nu\Delta\eta \hspace{1cm}\mbox{at $z=0$.}
\end{eqnarray}

It is well-known that the modulations of weakly nonlinear (undamped) gravity waves in deep water,
with basic wave number $k_0$ and frequency $\omega_0(k_0)$, can be described by the non-dissipative 
nonlinear Schr\"odinger (NLS) equation for the envelope $A$ of a Stokes wavetrain (written here in the 2D case)
\begin{equation}
i\frac{\partial A}{\partial t} -\frac{\omega_0}{8 k_0^2}\frac{\partial^2 A}{\partial x^2} - \frac{1}{2}\omega_0 k_0^2|A|^2 A = 0.
\end{equation}
Recall that $A$ is the envelope of the normal canonical variable $a$. The Fourier harmonics $a$ satisfy the following relation:
\begin{equation}\nonumber
a = \sqrt{\frac{\omega}{2 |k|}}\eta_k + i \sqrt{\frac{|k|}{2\omega}}\psi_k,
\end{equation}
where $\eta_k$ is the Fourier transform of the free-surface elevation and $\psi_k$ the Fourier transform of the velocity 
potential evaluated on the free surface, see \cite{ZFL1992}.
What happens if one tries to derive a similar equation from the boundary conditions (\ref{linear_kinematic_N_S_new}) 
and (\ref{linear_Bernoulli})? Dissipation then appears naturally in the NLS equation in the following way:
\begin{equation}\label{NLS_damped}
i\frac{\partial A}{\partial t} -\frac{\omega_0}{8 k_0^2}\frac{\partial^2 A}{\partial x^2} - \frac{1}{2}\omega_0 k_0^2|A|^2 A = 
-2i\nu k_0^2 A.
\end{equation}
It is important to recognize that equation (\ref{NLS_damped}) has been used by many authors but without convincing
physical explanation. Authors either simply refer to previous papers or argue that it is the simplest
way to include dissipation! Here we provide a clear derivation of the damped NLS equation. It is also important
to point out that if vorticity effects are not included in the kinematic boundary condition, then the resulting
NLS equation looks much more complicated.

\section{Discussion}

\cite{RFF1991} derived a system of three equations in the limit of small viscosity: one for the velocity potential, 
one for the surface elevation and one for the vortical component of the velocity. \cite{LH1992} simplified these equations.
\cite{D1994} and \cite{SVBM2002} applied these equations to particular problems.

Here we have shown that the vortical component of velocity can be excluded from the equations, thus leaving only two equations
for a quasi-potential flow. A small amount of vorticity has been incorporated in the kinematic boundary condition.
 
It should be mentioned that there are several publications where the same dissipation is used in both
Bernoulli's equation and in the kinematic boundary condition. Such a choice is justified when one is interested in the 
numerical integration of potential flow equations. 
Among them one can refer to \cite{BakerMeironOrszag1989}, \cite{DKZ2003}, \cite{DKZ2004}, \cite{ZKPD2005}. 
This type of dissipation was used not only for the simulation of gravity waves but also for the study of capillary wave turbulence 
by \cite{PZ1996}.
The addition of dissipation is sometimes used to satisfy the radiation condition. Another question of interest is the value
that one is supposed to use for the kinematic viscosity $\nu$. The molecular value $\nu=10^{-6}$ m$^2/$s is far too small
to be significant for normal waves. In fact it seems that values which are of the order of $10^{-4}$ or even $10^{-3}$ are
commonly used. Therefore it is the eddy viscosity rather than the molecular viscosity that ought to be used for practical
applications. 

Finally, the set of dissipative equations we derived can be extended to interfacial waves and more generally to multi-layer
configurations. 

\section*{Acknowledgments}

A.I. Dyachenko is grateful for support from Centre National de la Recherche Scientifique. 
This work was also supported by the US Army Corps of Engineers Grant W912HZ-05-P-0351, by ONR Grant 
N00014-03-1-0648, NSF Grant DMS 0404577, RFBR Grant
06-01-00665, the Program ``Fundamental Problems in Nonlinear
Dynamics''  from the RAS Presidium, and Grant ``Leading Scientific
Schools of Russia''. F. Dias thanks E.O. Tuck and L. Lazauskas for interesting discussions and references on the topic
of viscous damping and water waves, as well as the staff of the library of the University of Adelaide, Australia, for providing access
to the first edition of Lamb's book on hydrodynamics.

\end{document}